\documentclass[12pt]{article}

\begin{document}
\begin{center}

\hfill\vbox{\hbox{IPPP/04/61}
            \hbox{DTP/04/122}}

{\Large \bf
New application of the the large-$N_c$ expansion:\\
comparison of the Gottfried and Adler sum rules}

\vspace{1cm}

{\bf D.J. Broadhurst$^{1}$, A.L. Kataev$^{2}$ and C.J. Maxwell
$^{3}$}

$^{1}$ Department of Physics and Astronomy, Open University,
Milton Keynes MK7 6AA, UK \\E-mail: D.Broadhurst@open.ac.uk\\
$^{2}$ Institute for Nuclear Research of the Academy of Sciences of Russia,
117312, Moscow, Russia\\E-mail: Kataev@ms2.inr.ac.ru\\
$^{3}$ Institute for Particle Physics Phenomenology, Durham, UK
\\E-mail: C.J.Maxwell@durham.ac.uk

\end{center}
\begin{center}
{\bf ABSTRACT}
\end{center}
The Adler sum rule for deep inelastic neutrino
scattering
measures the isospin of the nucleon, and is hence exact.
In contrast the Gottfried
sum rule for charged lepton scattering does receive perturbative and
non-perturbative
corrections.
We show that at two-loop level the Gottfried sum rule is suppressed by
a factor $1/{N}_{c}^{2}$ relative to higher moments, and we conjecture that
this
suppression holds to all-orders, and also for higher-twist effects.
It is further noted
that the {\it differences} between radiative corrections for higher moments
of neutrino
and charged lepton deep inelastic scattering, are ${1}/{N}_{c}^{2}$
suppressed at two-loops,
and this is also conjectured to hold to all-orders.
The $1/{N}_{c}^{2}$ suppression of perturbative
corrections to the Gottfried sum rule makes it plausible that the
deviations from the parton model
value are dominated by a light quark flavour asymmetry in the nucleon sea.
This asymmetry indeed
persists as ${N}_{c}\rightarrow{\infty}$ as predicted in a
chiral-soliton model.
\noindent
\vspace*{0.1cm}

In this talk
we describe the results of the
recent work of Ref. \cite{r1}, in which we are led to a conjecture concerning
 the radiative QCD corrections (both perturbative and higher-twist)
to non-singlet
neutrino nucleon and charged nucleon Deep Inelastic Scattering (DIS).
Let us begin by considering the isospin
Adler sum rule, which is the first non-singlet moment for neutrino DIS.
This has the parton model expression
\begin{eqnarray}
{I}_{A}&\equiv&\int_{0}^{1}\frac{dx}{x}\left[{F}_{2}^{{\nu}p}(x,{Q}^{2})-{F}_{2}^{{\nu}n}(x,{Q}^{2})\right]
\nonumber \\
&=& 2\int_{0}^{1}{dx}\left(u(x)-d(x)-{\bar{u}}(x)+{\bar{d}}(x)\right)
\nonumber \\
&=&4{I}_{3}=2\;.
\end{eqnarray}
Since isospin is conserved this sum rule has the special feauture that
it is exact, and receives
no perturbative or non-perturbative QCD
corrections. This expectation of ${I}_{A}=2$ is consistent
with existing neutrino-nucleon DIS \cite{r2}, which show no significant $Q^2$ variation in the range $
2{\rm{GeV}}^{2}\leq{Q}^{2}\leq30{\rm{GeV}}^{2}$ and give
\begin{equation}
{I}_{A}^{\rm{exp}}=2.02\pm{0.40}\;.
\end{equation}
The corresponding sum rule for charged-lepton-nucleon DIS has the form
\begin{eqnarray}
{I}_{G}({Q}^{2})&=&\int_{0}^{1}\frac{dx}{x}\left[{F}_{2}^{lp}(x,{Q}^{2})
-{F}_{2}^{ln}(x,{Q}^{2})\right]
\nonumber \\
&=&\frac{1}{3}\int_{0}^{1}{dx}\left(u(x)-d(x)+{\bar{u}}(x)-{\bar{d}}(x)\right)
\nonumber \\
&=&\frac{1}{3}-\frac{2}{3}\int_{0}^{1}{dx}\left({\bar{d}}(x)-{\bar{u}}(x)\right)\;.
\end{eqnarray}
If the nucleon sea were flavour symmetric with ${\bar{u}}(x)={\bar{d}}(x)$
then one has the valence contribution to
the Gottfried sum rule ${I}_{G}^{v}=1/3$ only. This value strongly {\it disagrees} with the data
as analysed by the NMC collaboration \cite{r3}
which gave the following result
\begin{equation}
{I}_{G}^{\rm{exp}}({Q}^{2}=4\;{\rm{GeV}}^{2})=0.235\pm{0.026}\;.
\end{equation}
In contrast to the Adler sum rule the Gottfried sum rule is not exact and
will be
modified by both perturbative and non-perturbative corrections. The perturbative corrections
to O(${\alpha}_{s}^{2}$)
were analysed numerically in Ref. \cite{r4}
and were found to be small. They
cannot explain the discrepancy between the NMC data and
the naive expectation ${I}_{G}^{v}=1/3$.
A possible resolution is the existence of a light quark flavour asymmetry with ${\bar{u}}(x,{Q}^{2})<{\bar{d}}(x,{Q}^{2})$
over a significant $x$-range.\\

In the case of flavour symmetric sea
the perturbative QCD corrections to the Gottfried sum rule can be
written in the form
\begin{equation}
{I_G^{v}}({Q}^{2})=A({\alpha}_{s}){C}^{(l)}({\alpha}_{s})\;,
\end{equation}
where $A$ is the anomalous dimension contribution and ${C}^{(l)}$ is the
coefficient function.
\begin{equation}
{C}^{(l)}({\alpha}_{s})=
\frac{1}{3}\left[1+
\sum_{n=1}^{\infty}{C}_{n}^{(l)N=1}
{\left(\frac{\alpha_s}{\pi}\right)}^{n}\right]\;.
\end{equation}
The coefficient function receives no corrections at
O(${\alpha}_{s}$), and numerical integration of the two-loop results of
Van Neerven and Zijlstra \cite{r5} gave \cite{r4}
\begin{eqnarray}
{C}^{(l)N=1}_{2}=(3.695{C}_{F}^{2}-1.847{C}_{F}{C}_{A})\;,
\end{eqnarray}
where ${C}_{A}=N_c$ and ${C}_{F}=({N}_{c}^{2}-1)/2{N}_{c}$ are QCD Casimirs.
Combining
with the anomalous dimension part then yields for ${N}_{F}=3$ quark flavours
\begin{equation}
{I}_{G}^{v}(Q^2)=\frac{1}{3}
\bigg[1+0.0355\left(\frac{{\alpha}_{s}}{\pi}\right)
+\left(-0.853+
\frac{{\gamma}_{2}^{N=1}}{64{\beta}_{0}}\right)
{\left(\frac{{\alpha}_{s}}{\pi}\right)}^{2}\bigg]~~.
\end{equation}
Here ${\gamma}_{2}^{N=1}$ is the three-loop anomalous dimension coefficient
for e ${I}_{G}^{v}(Q^2)$,
which at the time of the calculation of \cite{r4} was unknown.
The relative one-loop anomalous dimension
coefficient is zero,
and at two-loops the result of calculations  \cite{r6} is:
\begin{equation}
{\gamma}_{1}^{N=1}=-4({C}_{F}^{2}-{C}_{F}{C}_{A}/2)
[13+8{\zeta}(3)-12{\zeta}(2)]\;.
\end{equation}
It is noteworthy that this is proportional to the typical non-planar colour factor
$({C}_{F}^{2}-{C}_{F}{C}_{A}/2)$, which is O$(1/{N}_{c}^{2})$ suppressed relative to the
individual weights ${C}_{F}^{2}$ and ${C}_{F}{C}_{A}$. For higher moments $N>1$ this
cancellation does not occur. The formulation of the conjecture started with
our observing that the numerically calculated two-loop coefficient of Eq.(7)
can be rewritten in the form
\begin{eqnarray}
{C}_{2}^{(l)N=1}&=&(3.695{C}_{F}^{2}-1.847{C}_{F}{C}_{A})
\nonumber \\
&=&3.695({C}_{F}^{2}-{C}_{F}{C}_{A}/2.0005)\;.
\end{eqnarray}
So to four sugnificant figures the non-planar colour factor is reproduced. This suggests
the conjecture that in fact the perturbative corrections to the Gottfried sum rule are
purely non-planar and are suppressed in the large-$N_c$ limit.\\

At two-loops one can show that this is indeed the case. The two-loop coefficient
${C}_{2}^{(l)N=1}$ can be defined through the $N=1$ Mellin moment of NS lepton-nucleon DIS
\begin{equation}
{C}_{2}^{(l)N}=3\int_{0}^{1}{dx}[{C}^{(2),(+)}(x,1)+{C}^{(2),(-)}(x,1)]{x}^{N-1}\;.
\end{equation}
The two-loop functions ${C}^{(2),(+)}$ and ${C}^{(2),(-)}$ have been computed by
van Neerven and Zijlstra \cite{r5}, and confirmed using another technique by Moch
and Vermaseren \cite{r7}. For neutrinoproduction DIS the corresponding moments involve
these {\it same} functions,
\begin{equation}
{C}_{2}^{(\nu)N}=\frac{1}{2}\int_{0}^{1}{dx}[{C}^{(2),(+)}(x,1)-{C}^{(2),(-)}(x,1)]{x}^{N-1}\;.
\end{equation}
The $N=1$ case corresponds to the Adler sum rule which has vanishing corrections, so that
${C}_{2}^{(\nu)N=1}=0$, and we can conclude that
\begin{equation}
\int_{0}^{1}{dx}{C}^{(2),(+)}(x,1)=\int_{0}^{1}{dx}{C}^{(2),(-)}(x,1)\;.
\end{equation}
We can then use this relation to eliminate ${C}^{(2),(+)}$ from ${C}_{2}^{(l)N=1}$ to
obtain
\begin{equation}
{C}_{2}^{(l)N=1}=2\times3\int_{0}^{1}{dx}{C}^{(2),(-)}(x,1)\;.
\end{equation}
Thus the Gottfried sum rule perturbative coefficients only involve the
${C}^{(2),(-)}$ function. One can check directly from the explicit results
of \cite{r5} that whilst ${C}^{(2),(+)}$ receives both planar and non-planar
contributions, ${C}^{(2),(-)}$ is explicitly proportional to the non-planar
factor $({C}_{F}^{2}-{C}_{F}{C}_{A}/2)$. Performing the ${C}^{(2),(-)}$ integration
to thirty significant figures using MAPLE, and matching to the expected structures
${\{}1,{\zeta}_{2},{\zeta}_{3},{\zeta}_{4}{\}}$, gives an analytical formula for
the two-loop coefficient,
\begin{equation}
{C}_{2}^{(l)N=1}=\left[-\frac{141}{32}+\frac{21}{4}{\zeta}(2)-\frac{45}{4}{\zeta}(3)+12{\zeta}(4)\right]
{C}_{F}({C}_{F}-{C}_{A}/2)\;.
\end{equation}
To finally show that to O(${\alpha}_{s}^{2}$) the perturbative corrections to the Gottfried
sum rule are suppressed in the large-${N}_{c}$ limit one needs to compute ${\gamma}_{2}^{N=1}$.
The recent calculation of three-loop non-singlet splitting functions by
Moch, Vermaseren and
Vogt \cite{r8} enabled us to compute this with the result
\begin{eqnarray}
\gamma_2^{N=1}&=&
(C_F^2-C_AC_F/2)\bigg\{
C_F\bigg[
290-248\zeta(2)
+656\zeta(3)
\nonumber
\\&&{}
-1488\zeta(4)+832\zeta(5)
+192\zeta(2)\zeta(3)
\bigg]
\nonumber
\\&&{}
+C_A\bigg[
{1081\over9}+{980\over3}\zeta(2)-
{12856\over9}\zeta(3)
\nonumber
\\&&{}
+{4232\over3}\zeta(4)
-448\zeta(5)
-192\zeta(2)\zeta(3)
\bigg]
\nonumber
\\&&{}
+N_F\bigg[
-{304\over9}-{176\over3}\zeta(2)+{1792\over9}\zeta(3)
-{272\over3}\zeta(4)\bigg]\bigg\}\\
\nonumber
&\approx&161.713785 - 2.429260\,N_F
\end{eqnarray}
which was obtained using the results of \cite{r9}.
There is indeed an overall non-planar colour factor.\\

One can extend the conjecture to higher moments $N>1$. ${C}^{(l)N}$
and ${C}^{(\nu)N}$ both contain the {\it same} ${C}^{(2),(+)}$ term,
and have an opposite sign ${C}^{(2),(-)}$ term. This immediately
implies that at two-loops the coefficient functions for higher moments
have identical planar contributions and {\it differ} by non-planar
contributions which are suppressed in the large-$N_c$ limit.
 The anomalous
dimension coefficients for general moments of non-singlet lepton-nucleon
and neutrino-nuleon DIS, ${\gamma}_{n}^{(l)N}$ and ${\gamma}_{n}^{(\nu){N}}$,
can be related to splitting functions ${P}^{(n)+}(x)$ and
${P}^{(n)-}(x)$, \cite{r6,r7}
\begin{equation}
{\gamma}_{n}^{(l)N}=-2\int_{0}^{1}{dx}{P}^{(n)+}(x){x}^{N-1}
\end{equation}
and
\begin{equation}
{\gamma}_{n}^{(\nu){N}}=-2\int_{0}^{1}{dx}{P}^{(n)-}(x){x}^{N-1}\;.
\end{equation}
At both two-loops \cite{r6} and at three-loops \cite{r7} one can
check that the difference of $(+)$ and $(-)$ splitting functions,
${P}^{(n)+}(x)-{P}^{(n)-}(x)$, is proportional to ${C}_{F}({C}_{F}-{C}_{A}/2)$
and is non-planar. This immediately demonstrates, using
Eqs.(17) and (18) above,
 that up to and including three-loop order the anomalous dimension coefficients
 ${\gamma}_{n}^{(l)N}$
and ${\gamma}_{n}^{(\nu){N}}$ {\it differ} by non-planar terms, supressed in
the large-${N}_{c}$ limit. Notice that the $N=1$ moment is again a special case
since the vanishing corrections to the Adler sum rule require that ${\gamma}_{n}^{(\nu){N=1}}=0$.
From Eq.(17) this implies that $\int_{0}^{1}{dx}{P}^{(n)-}(x)=0$, and hence given
that the difference of splitting functions is non-planar and for $n=1,2$
${\gamma}_{n}^{(l)N=1}$  in Eqs. (9),(16)
for the Gottfried sum rule is also non-planar.
 The conjecture is that these features persist
at all-loops. We can formulate the conjecture in a more precise way by
introducing the ``planar approximation'' \cite{r10}.\\

The {\it planar approximation} retains only those terms at O(${\alpha}_s^{n}$)
which contain the leading $N_c$ behaviour for each possible power of $N_F$. That is
we define
\begin{equation}
{C}_{n}^{N}{|}_{\rm{planar}}={C}_{F}\sum_{i=0}^{n-1}{{\cal{C}}}_{n,i}^{N}{N}_{F}^{n-1-i}{N}_{c}^{i}\;,
\end{equation}
where the ${{\cal{C}}}_{n,i}^{N}$ are pure numbers. The above conjectures then amount
to the statement that
\begin{equation}
{{\cal{C}}}_{n,i}^{(l)N}=6{{\cal{C}}}_{n,i}^{(\nu)N}\;,
\end{equation}
or equivalently that
\begin{equation}
{C}_{n}^{(l)N}{|}_{\rm{planar}}=6{C}_{n}^{(\nu)N}{|}_{\rm{planar}}\;,
\end{equation}
for all moments and for all-loops. The relative factor of $6$ simply reflects
the relative normalisation of the parton model sum rules. We should note
that the neutrino-nucleon moments will involve quark-loop terms involving ${N}_{F}{d}^{abc}{d}_{abc}/{N}_{c}$,
and we are assuming that such terms are discarded.
The $N=1$ case is
special precisely because the Adler sum rule is exact, which ensures that
${C}_{n}^{(l)N=1}{|}_{\rm{planar}}=0$ for the Gottfried sum rule.\\

If indeed the perturbative corrections to the Gottfried sum rule are non-planar
then this suggests that the infrared renormalons associated with its coefficient function
are also suppressed in the large-$N_c$ limit, but since IR renormalon ambiguities are
of the same form as OPE higher-twist power corrections, this implies that such higher-twist
corrections are also suppressed in the large-$N_c$ limit. This raises the possibility that
the dominant corrections to ${I}_{G}^{v}$ arise from a light quark flavour asymmetry. One
way of modelling this non-perturbative effect is the chiral soliton model,
which has been used in Ref. \cite{r11}
to estimate corrections
to the Gottfried sum rule. In this model one finds
\begin{equation}
\frac{1}{2}(3{I}_{G}^{v}-1)=\int_{0}^{1}{dx}\left({\bar{u}}(x)-{\bar{d}}(x)\right)=O({N}_{c}^{0}).
\end{equation}
So the flavour asymmetry {\it persists} in the large-$N_c$ limit. These authors obtained an
estimate consistent with $I_G$ in the range $0.219$ to $0.178$, in fair agreement with
${I}_{G}^{\rm{exp}}=0.235{\pm}0.026$.\\

Whilst we believe that there is compelling evidence that our general conjectures
are correct, what is clearly lacking is a proof. A further check will
be possible when the functions ${C}^{(3),(+)}(x)$ and ${C}^{(3),(-)}(x)$ are
known, work on this is underway at present \cite{r12}.
The conjecture could then
be confirmed up to and including three-loops.

{\bf Acknowledgements}

C.J.M. would like to thank the Organizers of the ICHEP 04 Conference
in Beijing for hospitality in China where
he presented the material described above.
A similar talk was presented
by A.L.K. at the 13th International Seminar Quarks2004, Pushkinskie Gory, Russia,
May 24-30, 2004. His work is supported by RFBR Grants N 03-02-17047,
03-02-17177.

\end{document}